\documentstyle[12pt]{article}

 \def\appendix{\par
 \setcounter{section}{0}
 \setcounter{subsection}{0}
 \def\thesection{\Alph{section}}
 \def\theequation{\thesection.\arabic{equation}}}
 \def\thebibliography#1{\subsection*{References}\list
 {[\arabic{enumi}]}{\settowidth\labelwidth{[#1]}
 \leftmargin\labelwidth
 \advance\leftmargin\labelsep
 \usecounter{enumi}}
 \def\newblock{\hskip .11em plus .33em minus .07em}
 \sloppy\clubpenalty4000\widowpenalty4000
 \sfcode`\.=1000\relax}

\def\a{\alpha}
\def\b{\beta}
\def\c{\chi}
\def\d{\delta}
\def\e{\epsilon}
\def\ve{\varepsilon}
\def\f{\phi}
\def\g{\gamma}
\def\h{\eta}
\def\j{\psi}
\def\k{\kappa}
\def\l{\lambda}
\def\m{\mu}
\def\n{\nu}

\def\q{\theta}
\def\r{\rho}
\def\s{\sigma}
\def\t{\tau}

\def\z{\zeta}

\def\F{\Phi}
\def\G{\Gamma}
\def\J{\Psi}
\def\Ld{\Lambda}

\def\S{\Sigma}

\def\inbar{\vrule height1.5ex width.4pt depth0pt}
\def\rlx{\relax\leavevmode}
\def\I{\leavevmode\hbox{\small1\kern-3.8pt\normalsize1}}
\def\openone{\leavevmode\hbox{\small1\kern-3.3pt\normalsize1}}
\def\Ione{\rlx{\rm 1\kern-2.7pt l}}
\font\cmss=cmss10
\font\cmsss=cmss10 at 7pt
\def\ZZ{\rlx\leavevmode
             \ifmmode\mathchoice
                    {\hbox{\cmss Z\kern-.4em Z}}
                    {\hbox{\cmss Z\kern-.4em Z}}
                    {\lower.9pt\hbox{\cmsss Z\kern-.36em Z}}
                    {\lower1.2pt\hbox{\cmsss Z\kern-.36em Z}}
               \else{\cmss Z\kern-.4em Z}\fi}
\def\Ik{\rlx{\rm I\kern-.18em k}}  
\def\IC{\rlx\leavevmode
             \ifmmode\mathchoice
                    {\hbox{\kern.33em\inbar\kern-.3em{\rm C}}}
                    {\hbox{\kern.33em\inbar\kern-.3em{\rm C}}}
                    {\hbox{\kern.28em\sinbar\kern-.25em{\rm C}}}
                    {\hbox{\kern.25em\ssinbar\kern-.22em{\rm C}}}
             \else{\hbox{\kern.3em\inbar\kern-.3em{\rm C}}}\fi}
\def\IP{\rlx{\rm I\kern-.18em P}}
\def\IR{\rlx{\rm I\kern-.18em R}}
\def\IN{\rlx{\rm I\kern-.20em N}}

\def\llsymbol#1{\@llsymbol{\@nameuse{c@#1}}}
\def\@llsymbol#1{\ifcase#1\or {}\or {'}\or {''}\or {'''}\or
   {''''}\or {'''''}\or  \else\@ctrerr\fi\relax}

\newcounter{contador}
\newcommand{\letra}{
   \stepcounter{equation}
   \setcounter{contador}{\value{equation}}
   \setcounter{equation}{0}
   \renewcommand{\theequation}{\thecontador.\alph{equation}}}
\newcommand{\antiletra}{
   \renewcommand{\theequation}{\arabic{equation}}
   \setcounter{equation}{\value{contador}}}

\def\acknowledgement{\if@twocolumn
\section*{Acknowledgements}
\else \normalsize
\begin{center}
{\bf Acknowledgements\vspace{-.5em}\vspace{0pt}}
\end{center}
\quotation
\fi}
\def\endacknowledgement{\if@twocolumn\else\endquotation\fi}

\newcommand{\ol}\overline
\newcommand{\ti}\tilde
\newcommand{\wt}\widetilde
\newcommand{\wh}\widehat
\newcommand{\bv}\breve
\newcommand{\dg}\dagger
\newcommand{\dr}{\stackrel{\mbox{\scriptsize DR}}\longrightarrow}

\newcommand{\sqed}{\mbox{\scriptsize SQED}}
\newcommand{\sd}{\mbox{\scriptsize SD}}
\newcommand{\C}{^{\mbox{\scriptsize c}}}
\newcommand{\sC}{\mbox{\scriptsize c}}

\newcommand{\QED}{QED$_{\mbox{\scriptsize 2+2}}$}
\newcommand{\Dddd}{$D$$=$$1$$+$$3$}
\newcommand{\DDdd}{$D$$=$$2$$+$$2$}
\newcommand{\Ddd}{$D$$=$$1$$+$$2$}

\newcommand{\AW}{{\mbox{\scriptsize AW}}}
\newcommand{\gf}{{\mbox{\scriptsize gf}}}
\newcommand{\inv}{{\mbox{\scriptsize inv}}}
\newcommand{\aand}{\;\;\;\mbox{and}\;\;\;}

\newcommand{\be}{\begin{equation}}
\newcommand{\ee}{\end{equation}}
\newcommand{\bl}{\begin{eqnarray}&}
\newcommand{\el}{&\end{eqnarray}}
\newcommand{\bq}{\begin{eqnarray}}
\newcommand{\eq}{\end{eqnarray}}
\newcommand{\0}{{\bf 0}}
\newcommand{\ts}{\textstyle}

\newcommand{\sx}{\sigma_x}
\newcommand{\sy}{\sigma_y}
\newcommand{\sz}{\sigma_z}
\newcommand{\sm}{{\s}^{\m}}

\newcommand{\smn}{\sigma^{\mu\nu}}
\newcommand{\ad}{{\dot\alpha}}
\newcommand{\bd}{{\dot\beta}}
\newcommand{\gd}{{\dot\gamma}}
\newcommand{\gm}{{\gamma}^m}

\newcommand{\upccad}{\widetilde\chi^{{\mbox{\scriptsize c}}\dot\alpha}}

\newcommand{\uppca}{\psi^{{\mbox{\scriptsize c}}\alpha}}

\newcommand{\upcad}{\widetilde\chi^{\dot\alpha}}

\newcommand{\uptad}{\widetilde\theta^{\dot\alpha}}

\newcommand{\qt}{\tilde\theta}
\newcommand{\qwt}{\widetilde\theta}
\newcommand{\ov}{\overline}
\newcommand{\pa}{\partial}

\def\sl#1{\rlap{\hbox{$\mskip 1 mu /$}}#1}	

\def\ssl#1{\rlap{\hbox{$ {\scriptstyle /}$}}#1}

\begin{document}

\title{\bf Super-${\tau}_{3}$QED and the dimensional reduction of $N$$=$$1$
super-{QED}$_{2+2}$ }

\author{{\it M. A. De Andrade}\thanks{Internet e-mail:
marco@cbpfsu1.cat.cbpf.br} ~and ~{\it O. M. Del Cima}{\thanks{Internet
e-mail: oswaldo@cbpfsu1.cat.cbpf.br}} \\
Centro Brasileiro de Pesquisas F\'\i sicas (CBPF) \\
Departamento de Teoria de Campos e Part\'\i culas (DCP)\\
Rua Dr. Xavier Sigaud, 150 - Urca \\
22290-180 - Rio de Janeiro - RJ - Brazil.}

\date{}

\maketitle

\begin{abstract}

In this work the supersymmetric gauge invariant action for the massive Abelian
$N$$=$$1$ super-{\QED} in the Atiyah-Ward space-time ({\DDdd}) is formulated.
The questions concerning the scheme of the gauge invariance in {\DDdd} by means
of gauging the massive $N$$=$$1$ super-{\QED} are investigated. We study how to
ensure the gauge invariance at the expenses of the introduction of a complex
vector superfield. We discuss the Wess-Zumino gauge and thereupon we conclude
that in this gauge, only the imaginary part of the complex vector field,
$B_{\m}$, gauges a $U(1)$-symmetry, whereas its real part gauges a Weyl
symmetry. We build up the gauge invariant massive term by introducing a pair of
chiral and anti-chiral superfields with opposite $U(1)$-charges. We carry out a
dimensional reduction {\it{\`a la}} Scherk of the massive $N$$=$$1$
super-{\QED} action from {\DDdd} to {\Ddd}. Truncations are needed in order to
suppress
non-physical modes and one ends up with a parity-preserving
$N$$=$$1$ super-QED$_{1+2}$ (rather than $N$$=$$2$) in $D$$=$$1$$+$$2$.
Finally, we show that
the $N$$=$$1$ super-QED$_{1+2}$ we have got is the supersymmetric version of
the ${\tau}_{3}$QED .

\end{abstract}

\newpage
\section{Introduction}

The idea of space-times with several time components and indefinite signature
has been taken seriously into account since a self-dual Yang-Mills theory in
4-dimensions {\cite{selfdym}} has been related to the Atiyah-Ward conjecture
{\cite{award}}. This theory is considered as a potential source for all
integrable models in lower dimensions, after some appropriate dimensional
reduction (DR) scheme is adopted.

Recently, it has been pointed out by Oogury and Vafa {\cite{ovafa}} that the
consistent backgrounds for $N$$=$$2$ string propagation correspond to self-dual
gravity (SDG) configurations in the case of closed $N$$=$$2$ strings, and
self-dual Yang-Mills (SDYM) configurations, coupled to gravity, in the case of
$N$$=$$2$ heterotic strings in four and lower dimensions. This result has been
reconfirmed by Gates and Nishino {\cite{gatesnish}} on the basis of
$\b$-function calculations for the Yang-Mills sector of the $N$$=$$2$ heterotic
string. More recently, Gates, Ketov and Nishino {\cite{gatesketnish1}} have
noticed the existence of Majorana-Weyl spinors in the Atiyah-Ward space-time,
{\em{i.e.}} {\DDdd}, and an $N$$=$$1$ self-dual supersymmetric Yang-Mills
(SDSYM) theory and a self-dual supergravity (SDSG) model were formulated for
the first time. Afterwards, an $N$$=$$2$ self-dual supersymmetric Yang-Mills
theory and $N$$=$$2$ and $N$$=$$4$ self-dual supergravities have been
formulated; in view of these results, it was also conjectured that the
$N$$=$$2$ superstrings have no possible counterterms at quantum level to all
orders in string loops {\cite{gatesketnish2}}.

The evidence for the close relationship between supersymmetric Chern-Simons
(SCS) theory and integrable models or topological theories gives enough
motivation to concentrate efforts in trying to understand more about field
theories in 3-dimensions. It has already been shown that the $N$$=$$1$ and
$N$$=$$2$ SCS theories in {\Ddd} are directly generated by the $N$$=$$1$ and
$N$$=$$2$ SDSYM theories in {\DDdd} by a suitable dimensional reduction and
truncation {\cite{drscs}}.

Since over the past years 3-dimensional field theories {\cite{djt}} have been
shown to play a central r\^ole in connection with the behaviour of
4-dimensional theories at finite temperature {\cite{12}}, as well as in the
description of a number of problems in Condensed Matter Physics
{\cite{13,domavro,qedtau3}}, it seems reasonable to devote some attention to
understand some peculiar features of gauge field dynamics in 3 dimensions.
Also, the recent result on the Landau gauge finiteness of Chern-Simons theories
is a remarkable property that makes 3-dimensional gauge theories so attractive
{\cite{finiteness}}. Very recently, this line of investigation has been
well-motivated in view of the possibilities of providing a gauge-theoretical
foundation for the description of Condensed Matter phenomena, such as
high-$T_{c}$ superconductivity {\cite{domavro}}, where the  QED$_{3}$ and
${\tau}_{3}$QED$_{3}$ {\cite{domavro,qedtau3}} are some of the theoretical
approaches that been forwarded as an attempt to understand more deeply about
high-$T_{c}$ materials.

The main purpose of this paper is to build up a superspace action that
describes a massive Abelian gauge model in {\DDdd}, namely, the $N$$=$$1$
version of {\QED}. Our work is organized as follows. In Section 2, we give the
details of the formulation of the $N$$=$$1$ supersymmetry in the Atiyah-Ward
space-time. The discussion and the explicit construction of an Abelian gauge
model with $N$$=$$1$ supersymmetry in {\DDdd} is the content of Section 3.
Here, we take massive matter fields, but the massless case is also contemplated
as a particular case of the former.

We show in Section 4 that, in carrying out a dimensional reduction {\it{\`a
la}} Scherk {\cite{scherk,sohnius}} of the massive $N$$=$$1$ super-{\QED}
action to {\Ddd} space-time dimensions, truncations are needed in order to
suppress non-physical modes and we end up with a parity-preserving $N$$=$$1$
super-${\tau}_{3}$QED (the supersymmetric version of the ${\tau}_{3}$QED)
{\cite{tau3}}, whose spectrum is free from tachyons and ghosts at tree-level.
Finally, in Section 5, we draw our general conclusions and present our
prospects for future work. Two appendices follow: the relevant aspects of
spinors in {\DDdd} and our notational conventions to work in {\DDdd} are listed
in the Appendix A. In the Appendix B, the conventions for {\Ddd} and some rules
for dimensional reduction are collected. The metric adopted throughout this
work for the Atiyah-Ward space-time is $\eta_{\m\n}=(+,-,-,+)$, $\m$,
$\n$=(0,1,2,3).

\section{$N\,$=1 supersymmetry and superfields in Atiyah-Ward space-time}

Ordinary space-time can be defined as the coset space (Poincar\'e
group)$/$(Lorentz group). Similarly, globally flat superspace can be defined as
the coset space (super-Poincar\'e group)$/$ (Lorentz group): its points are the
orbits which the Lorentz group sweeps out in the super-Poincar\'e group.
Elements of superspace are labeled by the Atiyah-Ward space-time coordinates;
$x^{\m}$, where $\m$$=$$(0,1,2,3)$, and the fermionic coordinates; $\q^{\a}$
and $\uptad$, where $\a$$=$$(1,2)$ and $\ad$$=$$(\dot 1,\dot 2)$. The fermionic
coordinates $\q$ and $\wt{\q}$ are Majorana-Weyl spinors.

Superfields are analytic functions of superspace coordinates, which should be
understood in terms of their power series expansions in $\q$ and $\wt{\q}$ with
coefficients which are themselves local fields over Minkowski space
{\cite{wessgris,piguet}}.

A compact and very useful technique for working out representations of the
supersymmetry algebra on fields was proposed by Salam and Strathdee
{\cite{Salam,Salam1}}: superfields in superspace. It is particularly useful for
$N$$=$$1$ theories, where their superfield structure is completely known. The
well-known algebra fulfilled by the generators of the supersymmetry in
$D$$=$$2$$+$$2$, $P_{\m}$, $Q_{\a}$ and ${\wt{Q}_{\ad}}$, is given by
\footnote{For notation and conventions in {\DDdd} see the Appendix A.}
\bq
&\{Q_{\a},{\wt{Q}_{\ad}}\}=2\;{\s}^{\m}_{\a
\ad}P_{\m}\;,\;\{Q_{\a},{{Q_{\b}}}\}=\{{\wt{Q}_{\ad}},{\wt{Q}_{\bd}}\}=0
\nonumber \\
&{\aand}[Q_{\a},P_{\m}]=[{\wt{Q}_{\ad}},P_{\m}]=0\;.\label{algebra}
\eq

The transformation law for a superfield, $F(x,\q,\wt{\q})$, is defined as
follows :
\be
\d F\equiv i\;(\ve\;{Q}+{\wt{\ve}}\;{\wt{Q}})F \label{susytransf}
\ee
where the parameters ${\ve}^{\a}$ and ${\wt{\ve}}^{\ad}$ are Majorana-Weyl
spinors as the same for the supercharges, $Q_{\a}$ and ${\wt{Q}_{\ad}}$, that
are given by
\be
Q_{\a}=-i(\pa_{\a}+i\sl{\pa}_{\a \ad}\uptad)\aand
{\wt{Q}_{\ad}}=-i(\wt{\pa}_{\ad}+i\sl{\wt{\pa}}_{\ad \a}\q^{\a})\;.
\label{charges}
\ee
The translations in superspace which result from the supersymmetry
transformations of the superfield $F(x,\q,\wt{\q})$ are presented below
\bq
&x^{\m}\longrightarrow{x^{\m}}+i\;{\ve}^{\a}{\s}^{\m}_{\a
\ad}{\uptad}+i\;{\wt{\ve}}^{\ad}{\wt{\s}}^{\m}_{\ad \a}\q^{\a}\;, \nonumber\\
&\q^{\a}\longrightarrow{\q^{\a}}+{\ve}^{\a}{\aand}{\uptad}
\longrightarrow{\uptad}+{\wt{\ve}}^{\ad}\;. \label{transl}
\eq

The covariant derivatives, $D_{\a}$ and ${{\wt{D}}_{\ad}}$, are such that the
application of them on a superfield, i.e., $D_{\a}F$ and ${{\wt{D}}_{\ad}}F$,
are covariant under supersymmetry transformations, this means that they are
also superfields, and these derivatives are such that
\be
D_{\a}=\pa_{\a}-i\sl{\pa}_{\a \ad}\uptad\aand
{\wt{D}_{\ad}}=\wt{\pa}_{\ad}-i\sl{\wt{\pa}}_{\ad \a}\q^{\a}\;,
\label{derivatives}
\ee
where they fulfil the following algebra
\bq
&\{D_{\a},{\wt{D}_{\ad}}\}=-2i\;{\s}^{\m}_{\a
\ad}\;{\pa}_{\m}\;,\;\{D_{\a},{{D_{\b}}}\}=\{{\wt{D}_{\ad}},{\wt{D}_{\bd}}\}=0
\nonumber \\
&{\aand}[D_{\a},{\pa}_{\m}]=[{\wt{D}_{\ad}},{\pa}_{\m}]=0\;.\label{dalgebra}
\eq

A chiral superfield, $\J$, is characterized by the covariant condition $\wt
D_\ad\J=0$; therefore, it follows that this superfield may be generally
parametrized as
\be
\J(x,\q,\wt{\q})=e^{i\qt\ssl{\tilde\pa}\q}\left[A(x)+i\q\j(x)+i\q^2F(x)
\right]\;, \label{chiral}
\ee
where $A$ is a complex scalar, $\j$ is a Weyl spinor and $F$ is a complex
scalar auxiliary field. The superfield, ${\J}^{\dg}$, that arises from the
Hermitian conjugation of the chiral superfield, ${\J}$,  is also a chiral
superfield (peculiarity of $D$$=$$2$$+$$2$), contrary to what happens in
$D$$=$$1$$+$$3$. The superfield ${\J}^{\dg}$ is given by
\be
{\J}^{\dg}(x,\q,\wt{\q})=e^{i\qt\ssl{\tilde\pa}\q}\left[A^*(x)+i\q\j\C(x)+
i\q^2F^*(x) \right]\;, \label{chiral+}
\ee
where we used the relations (\ref{conjcharge2}) and (\ref{conjcharge3}) of
Appendix A.

An anti-chiral superfield, ${\wt X}$, is such that it satisfies the constraint
$D_{\a}{\wt X}=0$, and may be written as follows
\be
{\wt X}(x,\q,\wt{\q})=e^{i\q\ssl{\pa}\qt}\left[B(x)+i\qwt\wt\c(x)+i\qwt^{2}G(x)
\right]\;,\label{antichiral}
\ee
where $B$ is a complex scalar, $\wt{\c}$ is a Weyl spinor and $G$ is a complex
scalar auxiliary field. Analogously to the previous case (by using
(\ref{conjcharge2}) and (\ref{conjcharge3})), the anti-chiral superfield ${{\wt
X}}^{\dg}$ is
\be
{\wt
X}^{\dg}(x,\q,\qwt)=e^{i\q\ssl{\pa}\qt}\left[B^*(x)+i\qwt\wt\c\C(x)+
i\qwt^{2}G^*(x) \right]\;.\label{antichiral+}
\ee

The rigid supersymmetry transformation law defined by eq.(\ref{susytransf})
yields for the components of ${\J}$ and ${{\wt X}}$ the following
transformations :
\bq
&\left\{\begin{array}{l}
\d A=i\ve^\a\j_\a \\
\d\j_\a=2\ve_\a F-2\wt\ve^\ad\wt\pa_{\ad\a}A \\
\d F=i\wt\ve^\ad\wt\pa_\ad^{\;\;\a}\j_\a
\end{array}\right. \aand
\left\{\begin{array}{l}
\d B=i\widetilde\ve^\ad\wt\c_\ad \\
\d\wt\c_\ad=2\widetilde\ve_\ad G-2\ve^\a\pa_{\a\ad}B \\
\d G=i\ve^\a\pa_\a^{\;\;\ad}\wt\c_\ad
\end{array}\right.\;\;\;\;\;.\label{scalarsmult}
\eq

Bearing in mind the necessity of the formulation of a supersymmetric gauge
theory in the Atiyah-Ward space-time, we are compelled to introduce a {\it
complex} vector superfield (a vector superfield without the reality
constraint),
$V$ :
\bq
V(x,\q,\qwt)\!\!\!&=&\!\!\!C(x)+i\q\z(x)+i\qwt\wt\h(x)+\frac12i\q^2M(x)+
\frac12i\qwt^2N(x)+\nonumber\\ &&+\frac12i\q\s^\m\qwt
B_\m(x)-\frac12\qwt^2\q\l(x)-\frac12\q^2\qwt\wt\r(x)-
\frac14\q^2\qwt^2 D(x)\;\;, \label{supervector}
\eq
where $C$, $M$, $N$ and $D$ are complex scalars, $\z$, $\wt\h$, $\l$ and
$\wt\r$ are Weyl spinors and $B_\m$ is a {\it complex} vector field. The
Hermitian conjugate, $V^{\dg}$, is given by
\bq
V^{\dg}(x,\q,\qwt)\!\!\!&=&\!\!\!C^*(x)+i\q\z\,\C(x)+i\qwt\wt\h\,\C(x)+
\frac12i\q^2M^*(x)+\frac12i\qwt^2N^*(x)+\nonumber\\ &&+\frac12i\q\s^\m\qwt
B^*_\m(x)-\frac12\qwt^2 \q\l\C(x)-\frac12\q^2 \qwt\wt\r\,\C(x)-
\frac14\q^2\qwt^2 D^*(x)\;\;, \label{supervector+}
\eq
where the relations (\ref{conjcharge2}) and (\ref{conjcharge3}) of Appendix A
have been used.

The field-strength superfields, $W_\a$ and ${\wt W}_\ad$ that satisfy
respectively, the chiral and anti-chiral conditions  $\wt{D}_{\bd}W_\a=0$ and
$D_{\b}{\wt W}_\ad=0$, are written as
\be
W_\a=\frac12{\wt D}^2D_\a V \aand
\wt{W}_\ad=\frac12{D}^2\wt{D}_\ad V \;\;;
\ee
in components we find
\bq
\left\{\begin{array}{l}
W_\a=e^{i\qt\ssl{\tilde\pa}\q}\left[\wh \l_\a+
\q^\b\left(\e_{\a\b}{\wh
D}-\s^{\m\n}_{\a\b}G_{\m\n}\right)+i\q^2\s^\m_{\a\ad}
\pa_\m{\wh{\wt\r}^{\,\ad}}\right] \\
\\
{\wt W}_\ad=e^{i\q\ssl{\pa}\qt}\left[\wh
{\wt\r}_\ad+\qwt^\bd\left(\wt\e_{\ad\bd}{\wh
D}-\wt\s^{\m\n}_{\ad\bd}G_{\m\n}\right)+
i\qwt^2\wt\s^\m_{\ad\a}
\pa_\m{\wh{\l}^{\,\a}}\right]
\end{array}\right.\;\;\;, \label{sstrenght}
\eq
where
\bq
\left\{\begin{array}{l}
\wh\l_\a=\l_\a-\s^{\m\;\;\ad}_{\;\,\a}\,\pa_\m\wt\h_\ad\\
\\
\wh D=D-\Box C\\
\\
{\wh{\wt\r}_\ad}=\wt\r_\ad-{\wt\s}^{\m\;\;\a}_{\;\,\ad}\,\pa_\m\z_\a\\
\\
G_{\m\n}=\pa_\m B_\n-\pa_\n B_\m
\end{array}\right.\;\;\;.
\eq

The supersymmetry transformation law (\ref{susytransf}) applied to $W_\a$ and
${\wt W}_\ad$ yields the following changes for its component
fields :
\be
\left\{
\begin{array}{l}
\d\wh\l_\a=\ve^\b\left(\e_{\a\b}\wh D-\s^{\m\n}_{\;\;\a\b}G_{\m\n}\right)\\
\\
\d\wh
D=i\ve^\a\s^{\m\;\;\ad}_{\;\,\a}\,\pa_\m\wh{\wt\r}_\ad+
i\wt\ve^\ad{\wt\s}^{\m\;\;\a}_{\;\,\ad}\,\pa_\m\wh\l_\a\\
\\
\d
G_{\m\n}=-i\ve^\a\s_{\m\a}^{\;\;\;\;\;\ad}\,\pa_\n\wh{\wt\r}_\ad+
i\wt\ve^\ad{\wt\s}_{\m\ad}^{\;\;\;\;\;\a}\,\pa_\n\wh\l_\a
-(\m\leftrightarrow\n)\\
\\
\d\wh{\wt\r}_\ad=\wt\ve^\bd\left(\wt\e_{\ad\bd}\wh
D-\wt\s^{\m\n}_{\;\;\ad\bd}G_{\m\n}\right)
\end{array}
\right.\;\;\;\;\;.
\ee

The charge conjugation of the field-strength superfields may be found as
\bq
\left\{\begin{array}{l}
W\C_\a=e^{i\qt\ssl{\tilde\pa}\q}\left[\wh \l\C_\a+
\q^\b\left(\e_{\a\b}{\wh
D}^*-\s^{\m\n}_{\a\b}G_{\m\n}^*\right)+i\q^2\s^\m_{\a\ad}
\pa_\m{\wh{{\wt\r}}^{\,\sC\ad}}\right] \\
\\
\wt{W}\C_\ad=e^{i\q\ssl{\pa}\qt}\left[\wh
{\wt\r}\,\C_\ad+\qwt^\bd\left(\wt\e_{\ad\bd}{\wh
D}^*-\wt\s^{\m\n}_{\ad\bd}G_{\m\n}^*\right)+
i\qwt^2\wt\s^\m_{\ad\a}
\pa_\m{\wh{{\l}}^{\,\sC\a}}\right]
\end{array}\right.\;\;\;, \label{sstrenghtc}
\eq
where the relations ({\ref{conjcharge1}}), ({\ref{conjcharge2}}) and
({\ref{conjcharge3}}) are used. The connection between the complex conjugation
and the charge conjugation of spinors plays an important r\^ole in the
formulation of the supersymmetric gauge invariant actions as will be seen in
the following sections.

\section{Massive Abelian $N\,$=1 super-QED$_{2+2}$}

The supersymmetric extension of the massive Abelian QED in {\Dddd} requires two
chiral superfields carrying opposite $U(1)$-charges {\cite{wesszugauge}}. On
the other hand, to introduce mass for the matter sector in {\DDdd}, without
breaking gauge-symmetry, we have to deal with four scalar superfields: a pair
of chiral and a pair of anti-chiral supermultiplets; the members of each pair
have opposite $U(1)$-charges.

The massive Abelian $N$$=$$1$ super-{\QED} is described by the action :
\footnote{In this paper we are adopting $ds$$\equiv$$d^4xd^2\q$,
$d\wt{s}$$\equiv$$d^4xd^2\qwt$ and $dv$$\equiv$$d^4xd^2{\q}d^2\qwt$ for the
superspace measures.}
\bq
S_{\inv}^{\AW}\!\!\!&=&\!\!\!-{1\over8}\left(\int{ds}\;W^{\sC} W +
\int{d\wt{s}}\;{\wt{W}}^{\sC} \wt{W}\right) +
\int{dv}\;\left(\J^{\dg}_+e^{4qV}{\wt X}_+ + \J^{\dg}_-e^{-4qV}{\wt X}
_-\right)+\nonumber\\&& +
\,i\,m\left(\int{ds}\; \J_+\J_--\int{d\wt{s}}\;{\wt X}_+ {\wt X}_-\right) +
\mbox{h.c.} \;\;\;\;\;, \label{massqed}
\eq
where $q$ is a dimensionless coupling constant and $m$ is a parameter with
dimension of mass. The $+$ and $-$ subscripts in the matter superfields refer
to their respective $U(1)$-charges. To build up the interaction terms, we have
used a mixing between the chiral and anti-chiral superfields (in order to
justify such a procedure, we refer to the works of Gates, Ketov and Nishino
{\cite{gatesketnish2}}). This mixed interaction term establishes that the
vector superfield must be {\it complex}.

The gauge-fixing action in superspace is given by :
\be
S_{\gf}^{\AW}=-{1\over4\a}\int{dv}\;\left({\wt D}^2 V^\dg \right)\left(D^2 V
\right) + \mbox{h.c.} \;\;\;\;\;, \label{supergf}
\ee
where $\a$ is the dimensionless gauge-fixing parameter. It is worthwhile to
specify that the superfield $V$ and its corresponding field-strength
superfields, $W$, $W\C$, $\wt W$ and $\wt W\C$, are the same as the ones given
in Section 2.

In the massive super-{\QED}-action, given by eq.(\ref{massqed}), the chiral
superfields $\J_+$ and $\J_-$ ($\wt D_\ad\J_{\pm}$$=$$0$), are defined as
follows :
\be
\J_\pm (x,\q,\wt{\q})=e^{i\qt\ssl{\tilde\pa}\q}\left[A_\pm (x)+i\q\j_\pm
(x)+i\q^2F_\pm (x) \right] \;\;, \label{psi+-}
\ee
where $A_\pm$ are complex scalars, $\j_\pm$ are Weyl spinors, and $F_\pm$ are
complex scalar auxiliary fields. Moreover, the anti-chiral superfields, ${\wt
X}_+$ and ${\wt X}_-$ ($D_\a \wt X_{\pm}$$=$$0$), are defined by :
\be
{\wt X}_\pm (x,\q,\wt{\q})=e^{i\q\ssl{\pa}\qt}\left[B_\pm (x)+i\qwt\wt\c_\pm
(x)+i\qwt^{2}G_\pm (x) \right] \;\;, \label{chi+-}
\ee
where $B_\pm$ are complex scalars, $\wt{\c}_\pm$ are Weyl spinors and, $G_\pm$
are complex scalar auxiliary fields.

The transformations of the scalar superfields, $\J_{\pm}$ and ${\wt X}_{\pm}$,
that ensures the gauge invariance of the massive super-{\QED}-action
(\ref{massqed}) are the following:
\be
\J_{\pm} \longrightarrow e^{\mp i4q\Lambda_{\pm}} \J_{\pm} \;,\;\;
\wt{D}_{\ad}\Lambda_{\pm}=0 \aand {\wt X}_{\pm} \longrightarrow e^{\mp
i4q\wt\Gamma_{\pm}} {\wt X_{\pm}} \;,\;\; D_{\a}\wt\Gamma_{\pm}=0 \;\;,
\label{gaugetrans}
\ee
with
\be
\Lambda_{\pm}(x,\q,\wt{\q})=e^{i\qt\ssl{\tilde\pa}\q}\left[{\Ld}_{1
\pm}(x)+i\q{\Ld}_{2 \pm}(x)+i\q^2{\Ld}_{3 \pm}(x) \right]
\label{lambdachiral}
\ee
and
\be
\wt\G_{\pm}(x,\q,\wt{\q})=e^{i\q\ssl{\pa}\qt}\left[\G_{1 \pm}(x)+i\qwt\wt\G_{2
\pm}(x)+i\qwt^{2}\G_{3 \pm}(x) \right]\;\;\;,
\label{gammantichiral}
\ee
where ${\Ld}_{1 \pm}$ and $\G_{1 \pm}$ are complex scalars, ${\Ld}_{2 \pm}$ and
$\wt\G_{2 \pm}$ are Weyl spinors and, ${\Ld}_{3 \pm}$ and $\G_{3 \pm}$ are
complex scalar auxiliary fields.

Taking into account the gauge invariance of the action (\ref{massqed}) and
assuming the transformations of the superfields $\J_{\pm}$ and ${\wt X}_{\pm}$
(\ref{gaugetrans}), it may be directly found that the superfield $V$ suffers
the following gauge transformations
\be
\d_{g}
V=i\left(\wt\G_+-\Ld_+^{\dg}\right)=i\left(\wt\G_--\Ld_-^{\dg}\right)\;\;,
\label{dgaugetrans}
\ee
which puts in evidence the necessity for a {\it complex} vector superfield,
$V$, in order to make possible the construction of a gauge-invariant action in
the Atiyah-Ward space-time.

In terms of the component fields, the gauge transformation (\ref{dgaugetrans}),
by considering the eqs. (\ref{supervector}), (\ref{lambdachiral}) and
(\ref{gammantichiral}), becomes
\bq
\left\{\begin{array}{l}
\d_{g} C=i\left(\G_{1 \pm}-\Ld_{1 \pm}^{*} \right)\\     \\
\d_{g} \z_{\a}=-i \Ld\C_{2 \pm\, \a}\;\;,\;\;\; \d_{g} \wt\h_{\ad}= i\wt\G_{2
\pm\, \ad} \\
\\
\d_{g} M=-2i{\Ld}^*_{3 \pm} \;\;,\;\;\; \d_{g} N=2i\G_{3 \pm} \\
\\
\d_{g} B_{\m}=2i \pa_{\m} (\G_{1 \pm}+\Ld_{1 \pm}^{*})\\
\\
\d_{g} \l_{\a}=-i\s^\m_{\a\ad}\pa_{\m}\wt\G_{2 \pm}^{\,\ad}\;\;,\;\;\; \d_{g}
{\wt\r}_{\ad}=i \wt\s^\m_{\ad\a}\pa_{\m} \Ld_{2 \pm}^{\sC\, \a}\\
\\
\d_{g} D=i \Box\left(\G_{1 \pm}-\Ld_{1 \pm}^{*} \right)
\end{array}\right.\;\;\;\;\;\;\;\;\;\;. \label{gaugecomp}
\eq
Meanwhile, assuming the Wess-Zumino gauge {\cite{Salam1,wesszugauge}}, by
fixing the following non-supersy\-mme\-tric gauge transformations :
\bq
\left\{\begin{array}{l}
\d_{g} C=-C=i\left(\G_{1 \pm}-\Ld_{1 \pm}^{*} \right)\\
\\
\d_{g} \z_{\a}=-\z_{\a}=-i \Ld\C_{2 \pm\, \a}\;\;,\;\;\; \d_{g} \wt\h_{\ad}=-
\wt\h_{\ad}= i\wt\G_{2 \pm\, \ad} \\
\\
\d_{g} M=-M=-2i{\Ld}^*_{3 \pm} \;\;,\;\;\; \d_{g} N=-N=2i\G_{3 \pm}
\end{array}\right.\;\;\;\;\;\;\;\;\;\;. \label{WZgauge}
\eq
By eliminating the compensating fields of the multiplet $V$, one is led to the
following remaining transformation in the Wess-Zumino gauge :
\be
\d_{g} B_{\m}= i~\pa_{\m}\b
\;\;\;, \label{compWZgauge}
\ee
where $\b$ is an arbitrary complex function. However, as we shall see below,
after analysing the complete action in the Wess-Zumino gauge, the matter-gauge
couplings indicate that indeed only the imaginary part of $B_{\m}$ displays the
transformation of a genuine gauge field, whereas the real part of
$B_{\m}$-field gauges a Weyl symmetry. At this point, it should be mentioned
that from now on, it will be omitted the hat ($\widehat{\;\;\;}$) symbol over
the components fields, $\l$, $D$ and $\wt\h$, since the calculations will
always be performed in the Wess-Zumino gauge.

Adopting the Wess-Zumino gauge, the following component-field action stems from
the superspace action of eq.(\ref{massqed}) :
\bq
S_{\inv}^{\AW}\!\!\!&=&\!\!\!\int{d^4x}\left\{
-\frac14i\right({\l\C}{\sl\pa}\wt\r+{\wt\r}\,\C{\wt{\sl\pa}}\l \left)-\frac18
G^{*}_{\m\n}G^{\m\n}-\frac14 D^*D +\right.
\nonumber\\
&&
-F_+^{*}G_+ - A_+^{*}\Box B_+ - {1\over2} i {\j _+\C} {\sl{\pa}} \wt{\c}_+ -
qB_{\m} \left({1\over2} i {\j_+ \C}   \sm \wt{\c}_+ + A_+^*{\pa}^{\m}B_+ -
B_+{\pa}^{\m}A_+^*  \right) + \nonumber \\
&&
+ iq\biggl(A_+^*\wt{\c}_+ {\wt{\r}} +B_+\j_+\C {\l} \biggr) -  \left( qD+
q^2B_{\m} B^{\m}\right)A_+^*B_+ +\nonumber\\
&&
-F_-^{*}G_- - A_-^{*}\Box B_- - {1\over2} i {\j _-\C} {\sl{\pa}} \wt{\c}_- +
qB_{\m} \left({1\over2} i {\j_- \C}   \sm \wt{\c}_- + A_-^*{\pa}^{\m}B_- -
B_-{\pa}^{\m}A_-^*  \right) + \nonumber \\
&&
- iq\biggl(A_-^*\wt{\c}_- {\wt{\r}} +B_-\j_-\C {\l} \biggr) +  \left(
qD-q^2B_{\m} B^{\m}\right)A_-^*B_- +
\nonumber\\
&&\left.
+m \biggl(\frac12i\j_+\j_- - \frac12i\wt\c_+\wt\c_-
-A_+F_--A_-F_++B_+G_-+B_-G_+ \biggr)
\right\}+\mbox{h.c.} \;\;\;\;\;. \label{massaction}
\eq
Also, in the Wess-Zumino gauge, the gauge-fixing action (\ref{supergf}) is
given in components by
\be
S_{\gf}^{\AW}={1\over 4\a}\int{d^4x}\left\{
-i\biggl({\l\C}{\sl\pa}\wt\r+{\wt\r}\,\C{\wt{\sl\pa}}\l \biggr)+\left(\pa^\m
B_{\m}^*\right)\left(\pa^\n B_{\n}\right)-D^*D \right\}+\mbox{h.c.} \;\;\;\;\;.
\label{supergfcomp}
\ee

Due to the fact that in massive super-{\QED} one must have two opposite
$U(1)$-charges to introduce mass at tree-level, and a complex vector superfield
in order to build up the gauge invariant interactions, we can read directly
from the action (\ref{massqed}), and the superfields, (\ref{supervector}),
(\ref{psi+-}) and (\ref{chi+-}), the following set of local
$U(1)_{\a}$$\times$$U(1)_{\g}$ transformations :
\bq
&\left\{\begin{array}{l}
\d_{g} {A}_{\pm}^*={\pm}i q\b(x) {A}_{\pm}^*\\
\\
\d_{g} \j^{\C}_{\pm}={\pm}i q\b(x) \j^{\C}_{\pm} \\
\\
\d_{g} {F}_{\pm}^*={\pm}i q\b(x) {F}_{\pm}^*
\end{array}\right. \aand
\left\{\begin{array}{l}
\d_{g} B_{\pm}={\mp}i q\b(x) B_{\pm}\\
\\
\d_{g} \wt\c_{\pm}={\mp}i q\b(x) \wt\c_{\pm} \\
\\
\d_{g} G_{\pm}={\mp}i q\b(x) G_{\pm}
\end{array}\right.\;\;\;\;\;\;\;\;\;\;, \label{U(1)+-sym}
\eq
where $\b$$\equiv$$\a$$-$$i\g$ is an arbitrary infinitesimal $C^\infty$ complex
function. Notice that the gauge transformations (\ref{U(1)+-sym}) read as above
because one has previously fixed to work in the Wess-Zumino gauge. As for the
gauge superfield components surviving the Wess-Zumino gauge, we have :
\bq
&\left\{\begin{array}{l}
\d_{g} \l=\d_{g} \wt\r=0\;\;\;,\\
\\
\d_{g} D=0 \;\;\;\;\mbox{and} \\
\\
\d_{g} B_{\m}= i\; \pa_{\m}\b \;\;\;.\label{gaugesupertrans+-}
\end{array}\right.
\eq

Therefore, in the Wess-Zumino gauge, the real part of $B_{\m}$ gauges the
$U(1)_{\g}$-symmetry with real gauge function $\g$, whereas its imaginary part
gauges the $U(1)_{\a}$-symmetry with real gauge function $\a$. The latter is an
ordinary phase symmetry, and we associate it with the electric charge. Indeed,
as we will see later on, the imaginary component of $B_{\m}$ will be taken as
the photon field. The parameter $\g$ generates a local Weyl-like invariance
{\cite{privcom}}. However, the vector field that gauges such a symmetry, namely
the real part of $B_{\m}$, will be suppressed in the process of dimensional
reduction, so that such an invariance will not leave track in \Ddd.

It should be emphasized that the mass bilinears in the action given by
eq.(\ref{massaction}) preserve the local
$U(1)_{\a}$$\times$$U(1)_{\g}$-symmetry, since their component matter fields
(fermions and scalars) carry opposite charges. Therefore, the opposite values
of the $U(1)$-charges play a central r\^ole in the process introducing mass for
the matter fields without breakdown of the gauge-symmetry, similarly to what
happens in {\Dddd}.

\section{$N\,$=1 super-$\tau_{3}$QED from Atiyah-Ward space-time}

It is well-known that outstanding supersymmetric models with extended
supersymmetry are closely related to simple ones in higher dimensions
{\cite{scherk,sohnius}}. As we are interested in simple supersymmetric models
in {\Ddd}, since these ones should be fruitful for applications in Condensed
Matter Physics {\cite{13}}, we propose here to investigate what kind of model
comes out after a suitable compactification is adopted to dimensionally reduce
Atiyah-Ward space-time to 3 space-time dimensions. Our propose is to carry out
a dimensional reduction{\footnote{One uses the trivial dimensional reduction
where the time-derivative, $\pa_3$, of all component fields vanishes,
$\pa_3{\cal F}$$=$$0$. See also the Appendix B.}} of the massive $N$$=$$1$
super-{\QED} {\it{\`a la}} Scherk {\cite{scherk}}. Bearing in mind that this
procedure should extend the supersymmetry {\cite{scherk,sohnius}} to
$N$$\,>$$1$, truncations will be needed in order to remain with a simple
supersymmetry and to suppress non-physical modes, {\em{i.e.}} spurious degrees
of
freedom coming from {\DDdd} dimensions.

To perform the dimensional reduction {\it{\`a la}} Scherk from {\DDdd} to
{\Ddd} of the action (\ref{massaction}), use has been made of the rules
presented in
the Appendix B (see (\ref{dr1})--(\ref{dr10})). As a result, it can be directly
found the following supersymmetric action in {\Ddd} :
\bq
S_{\inv}^{D=3}\!\!\!&=&\!\!\!\int{d^3\hat{x}}\left\{
-\frac14i\right({\ov\l}{\gm {\pa}_m}\r+{\ov\r}{{\gm {\pa}_m}}\l \left)-\frac18
\biggl(G^{*}_{mn}G^{mn}+2{{\pa}_m \f}^*{\pa}^m \f \biggr)-\frac14 D^*D +\right.
\nonumber\\
&&
-F_+^{*}G_+ - A_+^{*}\Box B_+ - {1\over2} i {\ov\j _+} {\gm {\pa}_m} {\c}_+ -
qB_{m} \left({1\over2} i {\ov\j_+ } \gm {\c}_+ + A_+^*{\pa}^{m}B_+ -
B_+{\pa}^{m}A_+^*  \right) + \nonumber \\
&&
+{1\over2}q \f {\ov\j_+ } {\c}_+ + q\biggl(A_+^* \ov{\c}_+\C {\r} -B_+\ov\j_+
{\l} \biggr) -  \left( qD+q^2B_{m} B^{m}+q^2\f^2\right)A_+^*B_+ +\nonumber \\
&&
-F_-^{*}G_- - A_-^{*}\Box B_- - {1\over2} i {\ov\j _-} {\gm {\pa}_m} {\c}_- +
qB_{m} \left({1\over2} i {\ov\j_- } \gm {\c}_- + A_-^*{\pa}^{m}B_- -
B_-{\pa}^{m}A_-^*  \right) + \nonumber \\
&&
-{1\over2} q\f {\ov\j_- } {\c}_- - q\biggl(A_-^* \ov{\c}_-\C {\r} -B_-\ov\j_-
{\l} \biggr) +  \left( qD-q^2B_{m} B^{m}-q^2\f^2\right)A_-^*B_- +\nonumber\\
&&\left.
-m \biggl(\frac12 \ov\j_+\C\j_- + \frac12\ov\c_+\C \c_-
+A_+F_-+A_-F_+-B_+G_--B_-G_+ \biggr)
\right\}+\mbox{h.c.} \;\;\;\;\;, \label{action3}
\eq
where, after dimensional reduction, the coupling constant $q$ has acquired
dimension of (mass)$^{1\over2}$. Furthermore, after performing the dimensional
reduction of the gauge-fixing (\ref{supergfcomp}), we found the following
gauge-fixing action in {\Ddd} :
\be
S_{\gf}^{D=3}={1\over 4\a}\int{d^3\hat{x}}\left\{
-i\biggl({\ov\l}{\gm {\pa}_m}\r+{\ov\r}{{\gm {\pa}_m}}\l \biggr)-\left(\pa^m
B^{*}_{m}\right)\left(\pa^n B_{n}\right)-D^*D \right\}+\mbox{h.c.} \;\;\;\;\;.
\label{action3gf}
\ee

Analysing the 3-dimensional action{\footnote{Note that, $\l$, $\r$, $\j$ and
$\c$ are now Dirac spinors in {\Ddd}.} given by eq.(\ref{action3}), it can be
easily shown that the spectrum will unavoidably be spoiled by the presence of
ghost fields, since the free sector of the action is totally off-diagonal.
Therefore, truncations are needed in order to remove the spurious degrees of
freedom, as well as to give rise to a simple supersymmetric action in {\Ddd}.
First of all, to make the truncations possible, we need to diagonalize the
whole free sector, in order that the ghost fields be identified.

In order to probe more deeply such a conclusion, we should diagonalize the
free gauge and matter sectors of the action (\ref{action3}). The
diagonalization is achieved by looking for suitable linear combinations of the
fields which yield a diagonal free action. After tedious
algebraic manipulations\footnote{For the calculations, use has been made of the
{\bf{MapleV}} software, since, to fix the appropriate parameters present in the
massive Abelian $N$$=$$1$ super-{\QED} action (\ref{massqed}) so as to get
(\ref{action3}), we had to invert and diagonalize $8$$\times$$8$ and
$16$$\times$$16$ matrices.}, we find the following transformations which
diagonalize the action $S_{\inv}^{D=3}$ :
\begin{enumerate}
\item{gauge sector :}
\be
\l=\frac1{\sqrt{2}} \left(\wh\r + \wh\l \right) \aand \r=\frac1{\sqrt{2}}
\left(\wh\r - \wh\l \right) \;\;\;;
\ee
\item{fermionic matter sector :}
\be
\j_\pm=\frac1{\sqrt2}\left(\wh\j_\pm \mp {\wh\j_\mp}^{\rm c}+\wh\c_\pm \pm
{\wh\c_\mp}^{\rm c}\right) \aand \c_\pm=\frac1{\sqrt2}\left(\wh\c_\pm \pm
{\wh\c_\mp}^{\rm c}-\wh\j_\pm \pm {\wh\j_\mp}^{\rm c}\right) \;\;\;;
\ee
\item{bosonic matter sector :}
\be
A_\pm=\frac1{\sqrt2}\left(\wh A_\pm -\wh B_\pm\right) \aand
B_\pm=\frac1{\sqrt2}\left(\wh A_\pm +\wh B_\pm\right)\;\;\;;
\ee
\be
F_\pm=\frac1{\sqrt2}\left(\wh F_\pm +\wh G_\pm\right) \aand
G_\pm=\frac1{\sqrt2}\left(\wh G_\pm -\wh F_\pm\right)\;\;\;.
\ee
\end{enumerate}
On the other hand, to simplify the Yukawa-interaction terms (gaugino-matter
couplings), we find that the following field redefinitions for the bosonic
matter sector are convenient :
\be
\wh A_\pm=\frac1{\sqrt2}\left(\bv A_\pm \mp\bv A_\mp^*\right) \aand \wh
F_\pm=\frac1{\sqrt2}\left(\bv F_\pm \mp\bv F_\mp^*\right)\;\;\;.
\ee

By replacing these field redefinitions into the action (\ref{action3}), one
ends up with a diagonalized action, where the fields, $\f$, $\wh\r$, $\wh\c_+$,
$\wh\c_-$, $\wh B_+$ and $\wh B_-$ appear like ghosts in the framework of an
$N$$=$$2$-supersymmetric model. Therefore, in order to suppress these
non-physical modes, truncations must be performed. Bearing in mind that we are
looking for an $N$$=$$1$ supersymmetric 3-dimensional model (in the Wess-Zumino
gauge), truncations have to be imposed on the ghost fields, $\f$, $\wh\r$,
$\wh\c_+$, $\wh\c_-$, $\wh B_+$ and $\wh B_-$. To keep $N$$=$$1$ supersymmetry
in the Wess-Zumino gauge, we must simultaneously truncate the component fields,
$\wh G_+$, $\wh G_-$, $D$, $a_m$ and $\t$ {\footnote{The $a_m$ field is the
real part of $B_m$, since we are assuming $B_m$$=$$a_m$$+$$iA_m$. Also, as
$\wh\l$ is a Dirac spinor, it can be written in terms of two Majorana spinors
in the following manner: $\wh\l$$=$$\t$$+$$i\bv\l$.}} . The truncation of $\t$
is dictated by the suppression of $a_m$. Now, the choice of truncating $a_m$,
instead of $A_m$, is based on the analysis of the couplings to the matter
sector: $A_m$ couples to both scalar and fermionic matter and we interpret it
as the photon field in 3 dimensions.

After these truncations are performed, and omitting the $(\widehat{\;\;\;})$
and $(\bv{\;\;\;})$ symbols, we find the following action in {\Ddd} :
\bq
S_{N=1}^{\t_3{\rm QED}}\!\!\!\!&=&\!\!\!\!\int{d^3\hat{x}}\left\{
{\frac 12}i{\ov\l}{\gm {\pa}_m}\l-\frac14 F_{mn}F^{mn} +\right.
\nonumber\\
&&
- A_+^{*}\Box A_+ - A_-^{*}\Box A_- + i {\ov\j _+} {\gm {\pa}_m} {\j}_+ + i
{\ov\j _-} {\gm {\pa}_m} {\j}_- + F_+^{*}F_+ + F_-^{*}F_- + \nonumber \\
&& -  q A_{m}\biggl({\ov\j_+ } \gm {\j}_+ - {\ov\j_- } \gm {\j}_- +
iA_+^*{\pa}^{m}A_+ - iA_-^*{\pa}^{m}A_- - iA_+{\pa}^{m}A_+^* +
iA_-{\pa}^{m}A_-^* \biggr) + \nonumber \\
&&
 - iq \biggl(A_+ \ov{\j}_+ {\l} - A_-\ov\j_- {\l} - A_+^{*} \ov{\l} {\j}_+ +
A_-^{*} \ov{\l} {\j}_- \biggr) +  q^2 A_{m} A^{m}\left(A_+^*A_+ + A_-^*A_-
\right) +\nonumber \\
&&\left.
- m \biggl( \ov\j_+\j_+ - \ov\j_- \j_- + A_+^{*}F_+ - A_-^{*}F_- + A_+F_+^{*} -
A_-F_-^{*} \biggr)
\right\} \;\;\;, \label{action3diag}
\eq
where it can be readily concluded that this is a supersymmetric extension of a
parity-preserving action, namely, ${\tau}_{3}$QED {\cite{qedtau3}}. However, to
render our claim more explicit, we are going to make use of a superspace
formulation, where the superfields are conveniently defined and the notational
conventions are fixed by the dimensional reduction. Before that, it should be
relevant to show how the gauge-fixing action (\ref{action3gf}) appears after
these suitable truncations :
\be
S_{\gf}^{\t_3{\rm QED}}={1\over 2\a}\int{d^3\hat{x}}\left\{
i{\ov\l}{\gm {\pa}_m}\l+\left(\pa^m A_m\right)^2 \right\} \;\;\;\;\;.
\label{action3gfdiag}
\ee

In order to formulate the $N$$=$$1$ super-${\tau}_{3}$QED action
(\ref{action3diag}) in terms of superfields, we refer to the work by Salam and
Strathdee {\cite{Salam}}, where the superspace and superfields in {\Dddd} were
formulated for the first time. Extending their ideas to our case in {\Ddd} (see
also ref.{\cite{wessgris}}), the
elements of superspace are labeled by $(x^m,\q)$, where $x^m$ are the
space-time coordinates and the fermionic coordinates, $\q$, are Majorana
spinors, $\q\C$$=$$\q$. {\footnote{The charge-conjugated spinor is defined by
$\j\C$$=$$-C\ol\j^T$, where $C$$=$$\sy$. The $\g$-matrices we are using arised
from the dimensional reduction to {\Ddd} are: $\g^m$$=$$(\sx,i\sy,-i\sz)$. Note
that for any spinorial objects, $\j$ and $\c$, the product $\ol\j\c$ denotes
$\ol\j_a \c_a$. For more details, see the Appendix B.}}

Now, we are ready to introduce the formulation of $N$$=$$1$
super-${\tau}_{3}$QED in terms of superfields. As a first step, we define the
complex scalar superfields with opposite $U(1)$-charges, $\F_+$ and $\F_-$ , as
\be
\F_\pm=A_\pm+\ol\q\j_\pm-\frac12\ol\q\q F_\pm \label{scalar3a} \aand
\F_\pm^\dg=A_\pm^*+\ol\j_\pm\q-\frac12\ol\q\q F_\pm^* \;\;\;\;,
\label{scalar3b}
\ee
where $A_\pm$ are complex scalars, $\j_\pm$ are Dirac spinors and $F_\pm$ are
complex scalar auxiliary fields.

In the Wess-Zumino gauge, the gauge superconnection, $\G_a$, is written as
\be
\G_a=i(\g^m\q)_aA_m+\ol\q\q\l_a \label{gauge3a} \aand
\ol{\G}_a=-i(\ol\q\g^m)_aA_m+\ol\q\q\ol\l_a \;\;\;\;, \label{gauge3b}
\ee
where $A_m$ is the gauge field and $\l_a$ is the gaugino (Majorana spinor).

Defining the field-strength superfield, $W_a$, according to :
\be
W_a=-\frac12{\ol D}_b D_a\G_b \;\;\;\;,
\ee
with superderivatives given by
\be
D_a={\ol\pa}_a-i(\g^m\q)_a\pa_m \aand {\ol
D}_a=-\pa_a+i(\ol\q\g^m)_a\pa_m~\;\;\;,
\ee
it can be found that
\letra
\bq
W_a\!\!\!&=&\!\!\!\l_a+\S^{mn}_{~~~ab}\q_bF_{mn}-
\frac{i}2\ol\q\q\;
\g^m_{~~ab}\left(\pa_m\l_b\right)  \label{strenght3a}\\
\noalign{\hbox{and}} \nonumber \\
\ol
W_a\!\!\!&=&\!\!\!\ol\l_a-\ol\q_b\S^{mn}_{~~~ba}F_{mn}+
\frac{i}2\ol\q\q\left(\pa_m\ol\l_b
\right)\g^m_{~~ba} \;\;\;\;, \label{strenght3b}
\eq
\antiletra
where $\S^{mn}$$=$${\ts\frac14}[\gamma^m,\gamma^n]$ are the generators of the
Lorentz group in {\Ddd}.

The gauge covariant derivatives we are defining for the matter superfields with
opposite $U(1)$-charges, $\F_+$ and $\F_-$, are given by
\be
\nabla_a\F_\pm=\left(D_a\mp iq\G_a\right)\F_\pm
\aand\ol\nabla_a\F^\dg_\pm=\left(\ol D_a\pm iq\ol\G_a\right)\F^\dg_\pm
\;\;\;\;, \label{deriv3}
\ee
where $q$ is a coupling constant with dimension of (mass)$^{1\over2}$.

By using the previous definitions of the superfields, (\ref{scalar3a}),
(\ref{gauge3a}), (\ref{strenght3a})-(\ref{strenght3b}), and the gauge covariant
derivatives, (\ref{deriv3}), we found how to build up the $N$$=$$1$
super-${\tau}_{3}$QED action, given by eq.(\ref{action3diag}), in superspace ;
it reads :
\be
S_{N=1}^{\t_3{\rm QED}}\!=\!\int\! d\hat v\left\{\;-{\frac12}\ol
WW+(\ol\nabla\F_+^\dg)(\nabla\F_+)+
(\ol\nabla\F_-^\dg)(\nabla\F_-)+
2m(\F_+^\dg\F_+-\F_-^\dg\F_-)\;\right\} \,,
\label{superqedtau3}
\ee
where the superspace measure we have adopted is $d\hat v$$\equiv$$d^3\hat x
d^2\q$ and the Berezin integral is taken as $\int\!d^2\q
$$=-$$\frac14\ol\pa\pa$ (see the Appendix B). Therefore, we finally show, by
using the superspace formulation (\ref{superqedtau3}), that the action
(\ref{action3diag}) we have found after a dimensional reduction {\it{\`a la}}
Scherk, and
some suitable truncations of the massive Abelian $N$$=$$1$ super-{\QED}, is
certainly the simple supersymmetric version of ${\tau}_{3}$QED.

The gauge-fixing action (\ref{action3gfdiag}) can be written in superspace as
\be
S_{\gf}^{\t_3{\rm QED}}\!=\!-{1\over 8\a}\int\! d\hat v\left\{\left({\ol D}
\G\right){\ol D} D \left({\ol D} \G\right)\right\}\;\;.
\label{supergftau3}
\ee

We conclude this section by pointing out that the massive Abelian $N$$=$$1$
super-{\QED} proposed in Section 3 shows interesting features, whenever an
appropriate dimensional reduction is performed. The dimensional reduction
{\it{\`a la}} Scherk we have applied to our problem becomes very attractive,
since, after doing some truncations to avoid non-physical modes, the $N$$=$$1$
super-${\tau}_{3}$QED is obtained as a final result. In fact, the Atiyah-Ward
space-time shows to be very fascinating as a starting point to formulate models
to be studied in lower dimensions.

\section{Discussions and general conclusions}

We attempted here to provide a connection between the $N$$=$$1$ supersymmetric
version of the parity-preserving ${\tau}_{3}$QED in {\Ddd} and the minimal
version of an $N$$=$$1$ supersymmetric {\QED} in Atiyah-Ward space-time.

The superspace formulation of the model in {\DDdd} reveals the peculiar
features of a complex gauge superconnection and an Abelian symmetry of the type
$U(1)_{\a}$$\times$$U(1)_{\g}$, where a local Weyl-like symmetry is present.
The reduction to {\Ddd} {\it{\`a la}} Scherk truncates, however, the gauge
field associated to this symmetry. But, we think that a better understanding of
such an invariance could be of relevance in connection with the formulation of
a conformally-invariant $N$$=$$1$ supergravity model coupled to the
super-{\QED} studied here {\cite{privcom1}}.

Space-time supersymmetry by itself is a good motivation to introduce self-dual
theories which have a good chance to be the generating theories for all
supersymmetric integrable models in lower dimensions. We would like to point
out that it would be worthwhile to study the possibility that our massive
Abelian model gives rise to a self-dual gauge field. This can be done on the
basis of the Parkes-Siegel formulation {\cite{parkessiegel}}, which after
carrying out a suitable dimensional reduction {\it{\`a la}} Nishino
{\cite{drscs}} to {\Ddd} generates a 3-dimensional model with a Chern-Simons
term for the $B_{\m}$-component of the vector supermultiplet {\cite{drdcp}}.

The Parkes-Siegel formulation for the massive Abelian $N$$=$$1$ super-{\QED}
coupled to a self-dual supermultiplet is achieved by introducing a chiral
multiplier superfield ($\wt D_\bd \Xi_{\a}=0$); its action is given by
\bq
S_{\sqed}^{\sd}\!\!\!&=&\!\!\!-\int{ds}\;\Xi^{\sC} W +
\int{dv}\;\left(\J^{\dg}_+e^{4qV}{\wt X}_+ + \J^{\dg}_-e^{-4qV}{\wt X}
_-\right)+\nonumber\\&& +
\,i\,m\left(\int{ds}\; \J_+\J_--\int{d\wt{s}}\;{\wt X}_+{\wt X}_-\right) +
\mbox{h.c.} \;\;\;\;\;, \label{masssdqed}
\eq
with
\be
\Xi_{\a}=e^{i\qt\ssl{\tilde\pa}\q}\left[A_{\a}+
{\q^\b}\left(\e_{\a\b}{E}-
\s^{\m\n}_{\a\b}H_{\m\n}\right)+
i\q^2{F}_{\a}\right]\;\;\;, \label{superspinor}
\ee
where $A_{\a}$ is a Weyl spinor, $E$ is a complex scalar, $H_{\m\n}$ is a
complex antisymmetric rank-2 tensor and ${F}_{\a}$ is a Weyl auxiliary spinor.

By adopting the Wess-Zumino gauge, the following component-field action stems
from the superspace action of eq.(\ref{masssdqed}) :
\bq
S_{\sqed}^{\sd}\!\!\!&=&\!\!\!\int{d^4x}\left\{-\frac12
H^{*}_{\m\n}\left(G^{\m\n}-\frac12 \epsilon^{\m\n\r\s}G_{\r\s} \right)
-i \biggl({A\C}{\sl\pa}\wt\r+F\C\l \biggr)- E^*D +\right.
\nonumber\\
&&
-F_+^{*}G_+ - A_+^{*}\Box B_+ - {1\over2} i {\j _+\C} {\sl{\pa}} \wt{\c}_+ -
qB_{\m} \left({1\over2} i {\j_+ \C}   \sm \wt{\c}_+ + A_+^*{\pa}^{\m}B_+ -
B_+{\pa}^{\m}A_+^*  \right) + \nonumber \\
&&
+ iq\biggl(A_+^*\wt{\c}_+ {\wt{\r}} +B_+\j_+\C {\l} \biggr) -  \left( qD+q^2
B_{\m} B^{\m}\right)A_+^*B_+ +\nonumber\\
&&
-F_-^{*}G_- - A_-^{*}\Box B_- - {1\over2} i {\j _-\C} {\sl{\pa}} \wt{\c}_- +
qB_{\m} \left({1\over2} i {\j_- \C}   \sm \wt{\c}_- + A_-^*{\pa}^{\m}B_- -
B_-{\pa}^{\m}A_-^*  \right) + \nonumber \\
&&
- iq\biggl(A_-^*\wt{\c}_- {\wt{\r}} +B_-\j_-\C {\l} \biggr) +  \left( qD- q^2
B_{\m} B^{\m}\right)A_-^*B_- +
\nonumber\\
&&\left.
+m \biggl(\frac12i\j_+\j_- - \frac12i\wt\c_+\wt\c_-
-A_+F_--A_-F_++B_+G_-+B_-G_+ \biggr)
\right\}+\mbox{h.c.}\;\;. \label{selfaction}
\eq
Therefore, it can be easily seen by action (\ref{selfaction}), that the field
equation of $H^*_{\m\n}$ gives the self-duality of the field-strength,
$G^{\m\n}$ :
\be
{\d S_{\sqed}^{\sd}\over \d H^*_{\m\n}}=0 \;\;\; \Longrightarrow \;\;\;
G^{\m\n}=\frac12 \epsilon^{\m\n\r\s}G_{\r\s}\;\;\;\;. \label{sdcond}
\ee

It would be a good suggestion to consider the dimensional reduction of the
action (\ref{selfaction}) according to the prescription proposed by Nishino in
ref.{\cite{drscs}}, and to try to understand the implications of the
3-dimensional theory in connection with the phenomenology of
${\tau}_{3}$QED$_{1+2}$. As a final remark, we point out that the
super-Yang-Mills version of the work presented here could be of interest in
association to the self-duality condition and the reduction of the model from
{\DDdd} to $D$$=$$1$$+$$1$, in order to check which sort of integrable model
may drop out from the reduction procedure.

\appendix
\section{General notations and conventions for
$D\,$=2+2}\setcounter{equation}{0}

We begin by reviewing some aspects of spinors living in the Atiyah-Ward
space-time. The Dirac spinor, $\J$, for even dimensions, may be represented, by
using the chiral operators, in terms of two Weyl spinors $\psi$ and $\wt{\c}$.
Each of the Weyl spinors transforms under the action of the group $SL(2,\IR)$
{\cite{gatesketnish1}}. In the Weyl representation the Dirac spinor takes the
form :
\bl
\J=
\left(\begin{array}{c}
\j\\
\wt\c\;
\end{array}\right)\;,                 \label{psi}
\el
where $\j$ and $\wt{\c}$ have the following components: $\j^\a$,
$\a$$=$$(1,2)$, and $\upcad$, $\ad$$=$$(\dot 1,\dot 2)$.

The Dirac $\gamma$-matrices can be represented by 4$\times$4 complex matrices
that satisfy the Clifford algebra :
\be
\{\g^\m,\g^\n\}=2\eta^{\m\n}\I_4\;\;\;,
\ee
where $\I_4$ is the 4$\times$4 identity matrix. Since the matrices
$-\g^{\m\dg}$, $\g^{\m*}$ and $-\g^{\m T}$ obey the same Clifford algebra as
the $\g^\m$, and there is only one irreducible representation of the Clifford
algebra by complex 4$\times$4 matrices up to equivalence transformations, there
exist matrices $A$, $B$ and $C$ with
\bq
\g^{\m\dg}\!\!\!&=&\!\!\!-A\g^\m A^{-1}\;\;\;,         \label{gmdag} \\
\g^{\m *}\!\!\!&=&\!\!\!B\g^\m B^{-1}\;\;\;,           \label{gmstar}\\
\g^{\m T}\!\!\!&=&\!\!\!-C\g^\m C^{-1}\;\;\;,     \label{gmtran}
\eq
where $A$=$\g^0\g^3$. The matrix, $B$, and the charge conjugation matrix, $C$,
in Weyl representation are given by
\be
B=
\left(\begin{array}{cc}
i\sz&\0\\
\0&i\sz
\end{array}\right)\aand                 
C=
\left(\begin{array}{cc}
\e&\0\\
\0&\wt\e
\end{array}\right)\;\;\;,                 \label{C}
\ee
where
\be
 \e=i\sy\aand\wt\e=-i\sy \;\;\;. \label{e}
\ee
The $\g$-matrices in this Weyl representation  are written as :
\be
\g^\m=
\left(\begin{array}{cc}
\0&\s^\m\\
\wt\s^\m&\0
\end{array}\right)\;\;\;, \label{gammamu}
\ee
\be
\g_5=\g^0\g^1\g^2\g^3=
\left(\begin{array}{cc}
\I_2&0\\
0&-\I_2
\end{array}\right)\;\;\;,
\ee
where $\I_2$ is the 2$\times$2 identity matrix. Besides, the $\s$-matrices of
(\ref{gammamu}) have the following components :
\bq
\s^\m\!\!\!&=&\!\!\!\left(-i\sx,\sy,-\sz,\I_2\right)\;\;\;,\label{smu}\\
\wt\s^\m\!\!\!&=&\!\!\!\left(i\sx,-\sy,\sz,\I_2\right)\;\;\;,\label{stmu}
\eq
where the usual Pauli matrices read
\be
\sx=\left(\begin{array}{cc}0&1\\
                           1&0
\end{array}\right)\;,\;\;
\sy=\left(\begin{array}{cc}0&-i\\
                           i&0
\end{array}\right)\;,\;\;
\sz=\left(\begin{array}{cc}1&0\\
                           0&-1
\end{array}\right)\;\;\;.
\ee
The $\ov{SO(2,2)}$-group has the following generators in the spinorial
representation :
\be
\Sigma^{\k\l}={\ts\frac14}[\g^\k,\g^\l]=
\left(\begin{array}{cc}
\s^{\k\l}&\0\\
\0&\wt\s^{\k\l}
\end{array}\right)\;\;\;. \label{generator}
\ee
Therefore, by using the eqs. (\ref{gammamu}) and (\ref{generator}), the $\s$
and $\wt\s$ matrices read
\be
\s^{\m\n}=\frac14(\s^\m\wt\s^\n-\s^\n\wt\s^\m) \aand
\wt\s^{\m\n}=\frac14(\wt\s^\m\s^\n-\wt\s^\n\s^\m)\;\;\;.
\ee
The complex conjugation of the matrices, $\sm$, $\wt\sm$, $\smn$ and $\wt\smn$
results \be
\s^{\m*}=\sz\s^\m\sz \aand \wt\s^{\m*}=\sz\wt\s^\m\sz \;\;\;;
\ee
\be
\s^{\m\n*}=\sz\s^{\m\n}\sz \aand \wt\s^{\m\n*}=\sz\wt\s^{\m\n}\sz\;\;\;.
\ee
Other useful relations involving the $\s$-matrices and their traces (Tr) used
in the calculations are given by :
\bq
{\mbox{Tr}}(\s^\m\wt\s^\n)\!\!\!&=&\!\!\!2\h^{\m\n}\;\;\;, \\
(\s^\m\wt\s^\n+\s^\n\wt\s^\m)^\a_{\;\;\b}\!\!\!&=
&\!\!\!2\h^{\m\n}\d^\a_{\;\;\b}\;\;\;,\\
(\wt \s^\m\s^\n+\wt
\s^\n\s^\m)^\a_{\;\;\b}\!\!\!&=&\!\!\!2\h^{\m\n}\d^\a_{\;\;\b}\;\;\;;
\eq
and
\bq
{\mbox{Tr}}(\s^{\m\n}\s^{\k\l})\!\!\!&=&
\!\!\!\frac12(\h^{\m\l}\h^{\n\k}-\h^{\m\k}\h^{\n\l}+\e^{\m\n\k\l})\;\;\;,\\
{\mbox{Tr}}(\wt\s^{\m\n}\wt\s^{\k\l})\!\!\!&=&
\!\!\!\frac12(\h^{\m\l}\h^{\n\k}-\h^{\m\k}\h^{\n\l}-\e^{\m\n\k\l})\;\;\;.
\eq

The charge-conjugated spinor, $\J\C$, is defined as follows
\be
\J\C=B\J^*=C\ol\J^T\;,              \label{chargespinor}
\ee
with $\ol{\J}$$=$$\J^{\dg}A$, where $A$$=$$\g^0\g^3$. In the Weyl
representation, the charge-conjugated spinor read as
\bl
\J\C=
\left(\begin{array}{c}
\j\C\\
\wt\c\C\;
\end{array}\right)\;,
\label{psic}
\el
where $\j\C$$\equiv$$i\sz\j^*$ and $\wt\c\C$$\equiv$$i\sz\wt\c^*$. For the
properties of charge-conjugated spinors living in $D$$=$$t$$+$$s$ space-time
dimensions see ref.{\cite{paperMarco}}.

The charge conjugation operation upon $\J$ for $D$$=$$2$$+$$2$ does not mix the
chiral sectors, since the matrix $B$ is diagonal in the Weyl representation.
Bearing in mind that the cove\-ring group of $SO(2,2)$ has the well-known
isomorphism $\ov{SO(2,2)}$$\cong$$SL(2,\IR)$$\otimes$$SL(2,\IR)$
{\cite{gilmore}}, it may be concluded that $\j\C$ and $\wt\c\C$ transforms in
the same manner as $\j$ and $\wt\c$ respectively, where the Weyl
conjugated-spinors $\j\C$ and $\wt\c\C$ have the following components: $\uppca$
and ${\upccad}$.

The Majorana spinor is defined by the constraint $\J$$=$$\J\C$. Therefore, due
to the fact that $B$ is diagonal in the Weyl representation, it follows
that in components we have $\j$$=$$\j\C$ and $\wt\c$$=$$\wt\c\C$. The Weyl
spinors which satisfy these constraints are called Majorana-Weyl spinors
{\cite{gatesketnish1}}.
In the case of $D$$=$$1$$+$$3$, it is well-known that
Majorana-Weyl spinors do not exist, since it is not possible to impose
simultaneously the Majorana and Weyl conditions.

\subsubsection*{Index conventions}

For all $\j$, $\wt\c$, $\s^\m$, $\wt\s^\m$, $\e$, $\wt\e$ that appear in the
text, we adopt the following conventions for the index structure :  $\j^\a$,
$\wt\c^\ad$, $\s^{\m\,\a}_{\;\;\;\;\;\ad}$, $\wt\s^{\m\,\ad}_{\;\;\;\;\;\a}$,
$\e_{\a\b}$, $\wt\e_{{\ad}{\bd}}\;$. In addition to this, we consider the
symbols $\e^{\a\b}$ and $\wt\e^{\ad\bd}$, such that
$\e^{\a\b}\e_{\b\g}$$=$$\d^{\a}_{\;\;\g}\;$ and
${\wt\e}^{\ad\bd}{\wt\e}_{\bd\gd}$$=$$\d^{\ad}_{\;\;\gd}\;$, which act on the
two independent $SL(2,\IR)$ sectors. Therefore, the spinor indices are raised
and lowered according to the rules :
\be
\j_{\a}=\e_{\a\b}\j^{\b}
\aand
\j^{\a}=\e^{\a\b}\j_{\b}\;\;\;;
\ee
\be
\wt\c_{\ad}=\wt\e_{\ad\bd}\wt\c^{\bd}
\aand
\wt\c^{\ad}=\wt\e^{\ad\bd}\wt\c_{\bd}\;\;\;.
\ee

The charge-conjugated Weyl spinors are given by
\be
\j^{\sC\,\a}=(i\sz\j^*)^{\,\a} \;\;,\;\;
\wt\c^{\sC\,\ad}=(i\sz\wt\c^*)^{\,\ad}\;\;\;. \label{conjcharge1}
\ee

Some of the $SL(2,\IR)$ invariant bilinears can be briefly written as
\be
\j^\a\j_\a\equiv\j^2 \aand \j^\a\l_\a\equiv\j\l=\l\j\;\;\;;
\ee
\be
\wt\c^\ad\wt\c_\ad\equiv{\wt\c}^2\aand
\wt\c^\ad\wt\r_\ad\equiv\wt\c\wt\r=\wt\r\wt\c\;\;\;.
\ee
Also, other bilinears are such that
\be
\j^\a\s^\m_{\a\ad}\wt\c^{\;\ad}\equiv\j\s^\m\wt\c=-\wt\c\wt\s^\m\j\;\;\;,
\ee
\be
\wt\r^{\,\ad}\wt\s^\m_{\ad\a}\l^{\a}\equiv\wt\r\wt\s^\m\l=-\l\s^\m\wt\r\;\;\;.
\ee
Their complex conjugation yields
\be
(i\j\l)^*=i\j\C\l\C \;\;,\;\; (i\wt\c\wt\r)^*=i\wt\c\C\wt\r\C \;\;\;;
\label{conjcharge2}
\ee
\be
(i\j\s^\m\wt\c)^*=i\j\C\s^\m\wt\c\C \;\;,\;\;
(i\wt\r\wt\s^\m\l)^*=i\wt\r\C\wt\s^\m\l\C \;\;\;. \label{conjcharge3}
\ee

Some useful relations for the Majorana-Weyl spinors $\q$ and $\wt\q$ follow :
\bq
\q^\a\q^\b\!\!\!&=&\!\!\!-\frac12\e^{\a\b}\q^2\;\;\;,\\
\q_\a\q_\b\!\!\!&=&\!\!\!\frac12\e_{\a\b}\q^2\;\;\;,\\
\wt\q^\ad\wt\q^\bd\!\!\!&=&\!\!\!-\frac12\wt\e^{\,\ad\bd}\wt\q^2\;\;\;,\\
\wt\q_\ad\wt\q_\bd\!\!\!&=&\!\!\!\frac12\wt\e_{\ad\bd}\wt\q^2\;\;\;,\\
\q\s^\m\wt\q\;\q\s^\n\wt\q\!\!\!&=&\!\!\!\frac12\q^2\wt\q^2\h^{\m\n}\;\;\;.
\eq

The fermionic derivatives are defined as :
\bq
\pa_\a\!\!\!&\equiv&\!\!\!\frac{\pa}{\pa\,\q^\a}\;\;\;,\\
\wt\pa_\ad\!\!\!&\equiv&\!\!\!\frac{\pa}{\pa\,\wt\q^\ad}\;\;\;.
\eq
Therefore, it follows that
\bq
\pa_\a\q^\b=\d_\a^{\,\,\,\b} &,\;\;&  \wt\pa_\ad\wt\q^\bd=\d_\ad^{\,\,\,\bd}
\;\;\;; \\
\pa^\a\q_\b=-\d^\a_{\,\,\,\b} &,\;\;& \wt\pa^\ad\wt\q_\bd=-\d^\ad_{\,\,\,\bd}
\;\;\;;\\
\pa_\a\q_\b=-\e_{\a\b} &,\;\;& \wt\pa_\ad\wt\q_\bd=-\wt\e_{\ad\bd} \;\;\;;\\
\pa^\a\q^\b=\e^{\a\b} &,\;\;& \wt\pa^\ad\wt\q^\bd=\wt\e^{\ad\bd} \;\;\;;\\
\pa_\a\q^2=2\q_\a  &,\;\;& \wt\pa_\ad\wt\q^2=2\wt\q_\ad \;\;\;;\\
\pa^\a\q^2=2\q^\a &,\;\;& \wt\pa^\ad\wt\q^2=2\wt\q^\ad \;\;\;; \\
\pa^2\q^2=-4 &,\;\;& \wt\pa^2\wt\q^2=-4 \;\;\;.
\eq

Throughout this work, the bosonic derivatives are defined by
\bq
\sl\pa \!\!\!&\equiv&\!\!\!\e\s^\m\pa_\m\;\;\;, \\
\sl\wt\pa \!\!\!&\equiv&\!\!\!\wt\e\wt\s^\m\pa_\m\;\;\;.
\eq

The superspace measures for the Atiyah-Ward space-time are
\be
ds \equiv d^4x d^2\q~~,~~~d\wt s \equiv d^4x d^2\wt\q \aand dv \equiv d^4x
d^2\q d^2\wt\q\;\;\;,
\ee
where the following normalization conditions are taken :
\be
\int d^2\q~\q^2 =1~~~{\rm and}~~\int d^2\wt\q~{\wt\q}^2=1\;\;\;.
\ee
For any superfield, $\Phi(x,\q, \wt\q)$, it can be directly shown that
\be
\int d^2\q\;\Phi=-\frac14 \partial^2\Phi=-\frac14 D^2\Phi\;|_{\q=\wt\q=0}~~,~~~
\int d^2\wt\q\;\Phi=-\frac14 \wt\partial^2\Phi=-\frac14 \wt
D^2\Phi\;|_{\q=\wt\q=0}
\ee
\be
\aand
\int d^2\q d^2\wt\q\;\Phi=\frac1{16} \partial^2 \wt\partial^2\Phi=\frac1{16}
D^2\wt D^2\Phi\;|_{\q=\wt\q=0}\;\;\;.
\ee

\section{General conventions for $D\,$=1+2 and some rules for dimensional
reduction}\setcounter{equation}{0}

In Section 4, we have adopted the metric $\eta_{mn}$$=$diag$(+,-,-)$, $m$,
$n$=(0,1,2), for {\Ddd}. The Dirac 2$\times$2 $\g$-matrices that satisfy the
Clifford algebra \be
\{\g^m,\g^n\}=2\eta^{mn}\I_2\;\;\;,
\ee
are as follows
\be
\g^m=i\s^m=-i\wt\s^m~=(\sx,i\sy,-i\sz)\;\;\;,
\ee
where $\s^m$ and $\wt\s^m$ are defined by eqs.(\ref{smu}) and (\ref{stmu}). In
{\Ddd}, the $\g$-matrices have an addition relation :
\be
\g^m\g^n=\h^{mn}\I_2+i\e^{mnl}\g_l\;\;\;.
\ee

We have the following generators of the $\ov{SO(1,2)}$-group in the spinor
representation :
\be
\Sigma^{kl}={\ts\frac14}[\gamma^k,\gamma^l]\;\;\;.
\ee
This yields an important relation used in the computation of the supergauge
sector action (\ref{superqedtau3}) :
\be
{\rm
Tr}\left(\Sigma^{kl}\Sigma^{mn}\right)=
\frac12\left(\h^{km}\h^{ln}-
\h^{kn}\h^{lm}\right)\;\;\;.
\ee
The charge conjugation matrix is found as
\be
C=-i\e=i\wt\e=\sy\;\;\;, \label{C3}
\ee
where $\e$ and $\wt\e$ are defined by eq.(\ref{e}).

The charge-conjugated and the adjoint spinors are defined by
\be
\j\C=-C{\ol\j}^T \aand \ol\j\equiv\j^\dg\g^0\;\;\;.\label{psiC3}
\ee

Some useful relations involving spinorial bilinears are listed below :
\bq
(\ol\j\c)^T\!\!\!&=&\!\!\!{\ol{\c}\C}\j\C \;\;\;,\nonumber\\
(i\ol\j\g^m\c)^T\!\!\!&=&\!\!\!-i{\ol{\c}\C}\g^m\j\C \;\;\;,\nonumber\\
(i\ol\j\g^m\pa_m\c)^T\!\!\!&=&\!\!\!i{\ol{\c}\C}\g^m\pa_m\j\C \;\;\;;\\
&&\nonumber\\
(\ol\j\c)^*\!\!\!&=&\!\!\!{\ol{\j}\C}\c\C \;\;\;,\nonumber\\
(i\ol\j\g^m\c)^*\!\!\!&=&\!\!\!i{\ol{\j}\C}\g^m\c\C \;\;\;,\nonumber\\
(i\ol\j\g^m\pa_m\c)^*\!\!\!&=&\!\!\!i{\ol{\j}\C}\g^m\pa_m\c\C\;\;\;;\\
&&\nonumber\\
(\ol\j\c)^\dg\!\!\!&=&\!\!\!\ol{\c}\j \;\;\;,\nonumber\\
(i\ol\j\g^m\c)^\dg\!\!\!&=&\!\!\!-i\ol{\c}\g^m\j \;\;\;,\nonumber\\
(i\ol\j\g^m\pa_m\c)^\dg\!\!\!&=&\!\!\!i\ol{\c}\g^m\pa_m\j\;\;\;.
\eq

As $\q$ is a Majorana spinor ($\q\C$$=$$\q$), by using the eqs.(\ref{C3}) and
(\ref{psiC3}), we found that
\bq
\q_a=(\ol\q C)_a \!\!\!&,&\!\! \ol\q_a=-(C\q)_a\;\;\;;\\
\q_a\q_b=-\frac12\ol\q\q C_{ab} \!\!\!&,&\!\! {\ol\q}_a{\ol\q}_b=\frac12\ol\q\q
C_{ab}\;\;\;;\\
\ol\q_a\q_b \!\!\!&=&\!\!\!\frac12\ol\q\q\;\d_{ab}\;\;\;, \\
\ol\q{\q} \!\!\!&\equiv&\!\!\!{\ol\q_a}\q_a \;\;\;.
\eq

The fermionic derivatives in {\Ddd} are defined as
\bq
\pa_a\q_b=\d_{ab}  \!\!&,&\!\!\! \ol\pa_a\ol\q_b=\d_{ab}\;\;\;;\\
  \pa_a\ol\q_b=C_{ab} \!\!&,&\!\!\!\ol\pa_a\q_b=C_{ab} \;\;\;;\\
         \pa_a\ol\q\q=-2{\ol\q}_a \!\!&,&\!\!\!{\ol\pa}_a\ol\q\q=2\q_a\;\;\;;\\
(\ol\pa\pa)(\ol\q\q)\!\!\!&=&\!\!\!-4 \;\;\;.
\eq

The superspace measure for {\Ddd} \footnote{The hat symbol $(\widehat{\;\;\;})$
over the 3-dimensional space-time coordinates is to distinguish from the
Atiyah-Ward ones.} space-time dimensions is
\be
d\hat v \equiv d^3\hat x d^2\q\;\;\;;
\ee
it respects the normalization condition
\be
\int d^2\q~\ol\q\q =1\;\;\;.
\ee

The covariant superderivatives satisfy the following algebra
\bq
\{D_a,D_b\}\!\!\!&=&\!\!\!2i(\g^m C)_{ab}\;\pa_m \;\;\;,\\
\{{\ol D}_a,{\ol D}_b\}\!\!\!&=&\!\!\!-2i(C\g^m)_{ab}\;\pa_m \;\;\;,\\
\{D_a,{\ol D}_b\}\!\!\!&=&\!\!\!2i\g^m_{ab}\;\pa_m\;\;\;,
\eq
where they are given by
\be
D_a={\ol\pa}_a-i(\g^m\q)_a\pa_m~,~{\ol D}_a=-\pa_a+i(\ol\q\g^m)_a\pa_m~\;\;\;.
\ee

For any superfield, $\Phi(\hat x,\q)$, it can be directly shown that
\be
\int d^2\q\;\Phi=-\frac14 {\ol\pa} \pa \Phi= -\frac14 {\ol D}
D\Phi\;|_{\q=0}\;\;\;.
\ee

The relation between the $\g$-matrices in \DDdd\ and in \Ddd\ are listed below,
\bq
\e\s^\m\!\!\!&=&\!\!\!(C\g^m , iC)\\
\wt\e\wt\s^\m\!\!\!&=&\!\!\!(C\g^m , -iC)\;\;\;,
\eq
where the left hand side is written in terms of the \DDdd\ quantities, whereas
the right side in terms of \Ddd\ ones.

To perform the dimensional reduction {\it{\`a la}} Scherk of the massive
$N$$=$$1$ super-{\QED}-action (\ref{massaction}) to {\Ddd}, use has been made
of the rules presented below. One uses the trivial dimensional reduction where
the time-derivative, $\pa_3$, of all component fields vanishes, $\pa_3{\cal
F}$$=$$0$. Also, it was assumed that $B_{\m}$ is reduced in the following
manner: $B^\m$$=$$(B^m , \f)$, where $\f$ is a complex scalar field. Note that
the Weyl spinors in {\DDdd} transform into Dirac ones after the dimensional
reduction is carried out to {\Ddd}. Now, we list the the following rules for
the dimensional reduction (DR) carried out ; they read :
\bq
G_{\m\n}^*G^{\m\n}&\dr& G_{mn}^*G^{mn}+2\pa_m\f^*\pa^m\f\;\;\;, \label{dr1}\\
\j\C\sl\pa\wt\c&\rightarrow&\ol\j\g^m\pa_m\c\;\;\;,\\
\wt\c\C\sl\wt\pa\j&\rightarrow&\ol\c\g^m\pa_m\j\;\;\;,\\
iB_\m\j\C\s^\m\wt\c&\rightarrow&iB_m\ol\j\g^m\c-\f\ol\j\c\;\;\;,\\
B_\m B\pa^\m A^*&\rightarrow&B_m B\pa^m A^*\;\;\;,\\
iA^*\wt\c\wt\r&\rightarrow&\;A^*{\ol{\c}\C}\r\;\;\;,\\
iB\j\C\l&\rightarrow&-\;B\ol\j\l\;\;\;,\\
B_\m B^\m A^*B&\rightarrow&B_m B^m A^*B+\f^2 A^*B\;\;\;,\\
i\j_+\j_-&\rightarrow&-\;{\ol{\j}_+\C}\j_-\;\;\;,\\
i\wt\c_+\wt\c_-&\rightarrow&\;{\ol{\c}_+\C}\c_-\;\;\;, \label{dr10}
\eq
where the fields in the left hand side are fields living in {\DDdd} and in the
other side appear fields living in {\Ddd}.

\small

\section*{Acknowledgements}

The authors are deeply indebted to Dr. J.A. Helay\"el-Neto for suggesting the
problem, for all the exhaustive discussions and the careful reading of the
manuscript. Dr. O. Piguet,  Dr. S.P. Sorella and Dr. L.P. Colatto are
acknowledged
for helpful discussions; Mrs. M.N.P. Magalh\~aes is acknowledged for
participation at
an early stage of this work. Thanks are also due to our colleagues at DCP,
in special to Dr.
S.A. Dias, for encouragement. CNPq-Brazil is acknowledged for invaluable
financial
help. Finally, the authors acknowledge the Organizing Committee of the
{\it Spring School
and Workshop on String Theory, Gauge Theory and Quantum Gravity '95 }
held at the
{\it International Centre for Theoretical Physics (ICTP)}.

\end{document}